\documentclass[]{spie}  

\usepackage{url}
\usepackage{wrapfig,framed}

\usepackage{amsmath,amsfonts,amssymb}
\usepackage{graphicx}
\usepackage[colorlinks=true, allcolors=blue]{hyperref}

\title{FIREBall-2: advancing TRL while doing proof-of-concept astrophysics on a suborbital platform}

\author[a]{Erika T. Hamden}
\author[b]{Keri Hoadley}
\author[b]{D. Christopher Martin}
\author[c]{David Schiminovich}
\author[d]{Bruno Milliard}
\author[e]{Shouleh Nikzad}
\author[d]{Ramona Augustin}
\author[d]{Philippe Balard}
\author[d]{Patrick Blanchard}
\author[f]{Nicolas Bray}
\author[b]{Marty Crabill}
\author[f]{Jean Evrard}
\author[f]{Albert Gomes}
\author[d]{Robert Grange}
\author[c]{Julia Gross}
\author[e]{April Jewell}
\author[b,e]{Gillian Kyne}
\author[g]{Michele Limon}
\author[b]{Nicole Lingner}
\author[b]{Mateusz Matuszewski}
\author[c]{Nicole Melso}
\author[f]{Frederi Mirc}
\author[f]{Johan Montel}
\author[c]{Hwei Ru Ong}
\author[b]{Donal O'Sullivan}
\author[d]{Sandrine Pascal}
\author[f]{Etienne Perot}
\author[d]{Vincent Picouet}
\author[f]{Muriel Saccoccio}
\author[c]{Brian Smiley}
\author[f]{Xavier Soors}
\author[f]{Pierre Tapie}
\author[d]{Didier Vibert}
\author[f]{Isabelle Zenone}
\author[c]{Jose Zorilla}

\affil[a]{University of Arizona, Steward Observatory, 933 N Cherry Ave, Tucson, AZ 85721, USA}
\affil[b]{California Institute of Technology, Division of Physics, Math, and Astronomy, 1200 E California Blvd, MC 278-17, Pasadena, CA 91105, USA}
\affil[c]{Columbia University, 550 W 120th St, New York, NY 10027, USA}
\affil[d]{Laboratoire d'Astrophysique de Marseille, 38 Rue Frédéric Joliot Curie, 13013 Marseille, France}
\affil[e]{NASA's Jet Propulsion Laboratory, 4800 Oak Grove Drive, Pasadena, CA 91109, USA}
\affil[f]{Centre national d'\'etudes spatiales, 18 Avenue Edouard Belin, 31400 Toulouse, France}
\affil[g]{Department of Physics and Astronomy, University of Pennsylvania, 209 South 33rd Street, Philadelphia, PA 19104, USA}

\authorinfo{Further author information: (Send correspondence to E.T.Hamden)\\E.T.Hamden: E-mail: hamden@email.arizona.edu}

\begin{document} 
\maketitle

\begin{abstract}

Here we discuss advances in UV technology over the last decade, with an emphasis on photon counting, low noise, high efficiency detectors in sub-orbital programs. We focus on the use of innovative UV detectors in a NASA astrophysics balloon telescope, FIREBall-2, which successfully flew in the Fall of 2018. The FIREBall-2 telescope is designed to make observations of distant galaxies to understand more about how they evolve by looking for diffuse hydrogen in the galactic halo. The payload utilizes a 1.0-meter class telescope with an ultraviolet multi-object spectrograph and is a joint collaboration between Caltech, JPL, LAM, CNES, Columbia, the University of Arizona, and NASA. The improved detector technology that was tested on FIREBall-2 can be applied to any UV mission. We discuss the results of the flight and detector performance. We will also discuss the utility of sub-orbital platforms (both balloon payloads and rockets) for testing new technologies and proof-of-concept scientific ideas.

\end{abstract}


\section{INTRODUCTION}
\label{sec:intro}  

The development of new detector technology is a key component of NASA's mission to advance our understanding of the universe around us. New advances in technology are always the precursors to discovery. Typically, NASA funds astrophysics technology development through the Astrophysics Research and Analysis (APRA) and Strategic Astrophysics Technology (SAT) programs, which are designed to take technology concepts from a low technology readiness level (TRL), up to a TRL of 6, at which point the technology can be included in proposals for space missions and are considered relatively low risk. 

In addition to funding technology development, NASA funds sub-orbital science investigations under APRA, either using sounding rockets or balloon flights to provide access to a near-space environment. These sub-orbital science investigations are typically seen as both interesting in their own right, but also as pathfinders for future space missions. The ability to increase a technology's TRL is just as important as the science goals of the missions. Because of the low cost of the sub-orbital program, riskier science can also be tested out. Additionally, these programs are not required to follow the more standard NASA management/risk mitigation program of a higher cost/profile mission, and so offer an opportunity for student and post-doc training that is not possible for Class-D or higher mission profiles. The ability to try out risky, interesting science, to test new technology, and to train future PIs is unique in NASA's portfolio. The sub-orbital program is a crucial, but under-appreciated, gem of NASA. 

Funded via the APRA program, the Faint Intergalactic-medium Redshifted Emission Balloon (FIREBall), is a perfect example of the benefits of building a sub-orbital program. The science is at the forefront of modern extra-galactic astronomy. The experiment has provided one for the first test flights of an important UV detector technology. The program as a whole has trained over 10 graduate students, with several becoming PIs in their own right. 

FIREBall is designed to discover and map faint emission from the circumgalactic medium of low redshift galaxies (0.3$<$z$<$1.0). This payload is a upgrade of FIREBall-1 \cite{2008Tuttle} (FB-1), a path-finding mission built by our team with two successful flights (2007 Engineering, 2009 Science). FB-1 provided the strongest constrains on intergalactic and circumgalactic (IGM, CGM) emission available from any instrument at the time \cite{2010Milliard}. FIREBall-2 significantly upgraded the spectrograph to increase the field of view, the number of targets observed, and most importantly, the overall system efficiency. The key component of FIREBall-2 was the use of a UV optimized, delta-doped, anti-reflection coated, electron multiplying CCD detector (EMCCDs), which has eight times better throughput that the FB-1 detector (a GALEX spare near-UV microchannel plate). FIREBall-2 had its first flight in the fall of 2018 from Fort Sumner, NM and acted as a pathfinder and test-bed for this new UV detector technology.

\subsection{FIREBall Science Goals}
The IGM is a fundamental element of our universe, as essential to our conception of the cosmos as stars and galaxies. The IGM delineates the large-scale structure of the universe and the majority of baryons at both high and low redshift reside here. It is the original source of fuel for star formation and is, in turn, altered by feedback from supernovae and galactic activity. Observing and understanding the IGM and its many parts are essential to our understanding of the universe as a whole. Yet for such a key part of galactic evolution, we know surprisingly little about the IGM, either within galactic halos as the CGM or part of the large-scale cosmic web.

Historically, observations of any parts of the IGM, even the brighter CGM material in-flowing into galaxies, were impossible to conduct given the low luminosity. Absorption line studies of the Ly-$\alpha$ forest are the only way we are certain it even exists in the low-redshift universe \cite{Croft+02, Tumlinson+13, Lee+14}. Observations of the brightest components of the IGM are at the very forefront of modern astronomy, requiring better technology, better analysis, and better observing techniques. Improved detectors, higher throughput optics, and state of the art instruments are right now opening a new field of IGM astronomy at high redshifts, where Ly-$\alpha$ is observable at visible wavelengths, as recent results from the Palomar and Keck Cosmic Web Imagers (PCWI \& KCWI \cite{Martin+14a, Martin+14b, Martin+15, Martin+16}) and Multi Unit Spectroscopic Explorer (MUSE \cite{Borisova+16, AB+16, AB+19}) attest.

\subsection{UV detector technology}

The history of UV astrophysics has been one of technology limitations. Before the invention and use of microchannel plates (MCPs) on UV missions, most missions used a variety of techniques to achieve limited efficiency in the UV. These techniques included single pixel photon counters, CCDs with a phosphorus overcoat, UV light flashes (prone to hysteresis), and scintillators. In the last few decades, MCPs have been used on many successful UV astrophysics missions, including EUV, FUSE, GALEX, and the Cosmic Origins Spectrograph (COS) on HST, \cite{1990Welsh,2000Moos,2005Morrissey,2012Green} as well as planetary missions. MCPs have been a reliable stalwart where previous technology failed. They provided better stability and throughput than previous technologies, and they truly shine in their extremely low noise properties, a benefit in the UV which has very low backgrounds \cite{2017Siegmund}. These detectors have been workhorses for the UV. Despite this, they  have several drawbacks: low efficiencies, especially compared to typical efficiencies of detectors at other wavelengths, count rate limits that exclude the brightest targets, and loss of gain (gain sag) over time, all of which have operational impacts \cite{2000Doliber,2018Oliveira,2018Sahnow}.

The recent development \cite{2012Nikzad, 2012Hamden} and flight testing of delta-doped silicon based CCD detectors has now opened a new opportunity in UV space astrophysics. Delta-doped CCDs in many ways are a perfect compliment to MCPs: delta-doped CCD efficiency, stability, and dynamic range are significantly better \cite{Nikzad+17}. MCPs will continue to play a role in future missions, but the high dynamic range of CCDs means they can make observations previously impossible with an MCP. The higher efficiency means that we will be able to observe more and fainter targets throughout the lifetime of a mission. 

Normal architecture CCDs that have been delta-doped do have additional noise sources that MCPs don’t and, like any other detector technology, will play an important role in missions where used appropriately. But the high sensitivity that results from the delta-doping process, when combined with electron multiplication technology of EMCCDs, provides a higher throughput detector with similar low noise properties to MCPs, which is ideal for use in the very low background environment of the UV. More detail on the EMCCD architecture is described in Section \ref{sec:EMCCD}. 

Future advances in UV detectors will likely include Magnetic Kinetic Inductance Detectors (MKIDS), which provide a low resolution energy measurement in addition to very high QE and limited noise, although their cryogenic operation is a challenge for spacecraft lifetimes and operations \cite{2012Mazin}. Skipper CCDs, once delta-doped to provide UV efficiency, could supplant EMCCDs for photon counting and low light level applications \cite{2017Tiffenberg}. The future of UV detectors, tested via the sub-orbital program, is full of possibilities and the technology breakthroughs of the last ten years will help spur 
UV missions under all parts of NASA's Science Mission Directorate.

\begin{figure}[t]
\centering
    \includegraphics[width=0.75\textwidth]{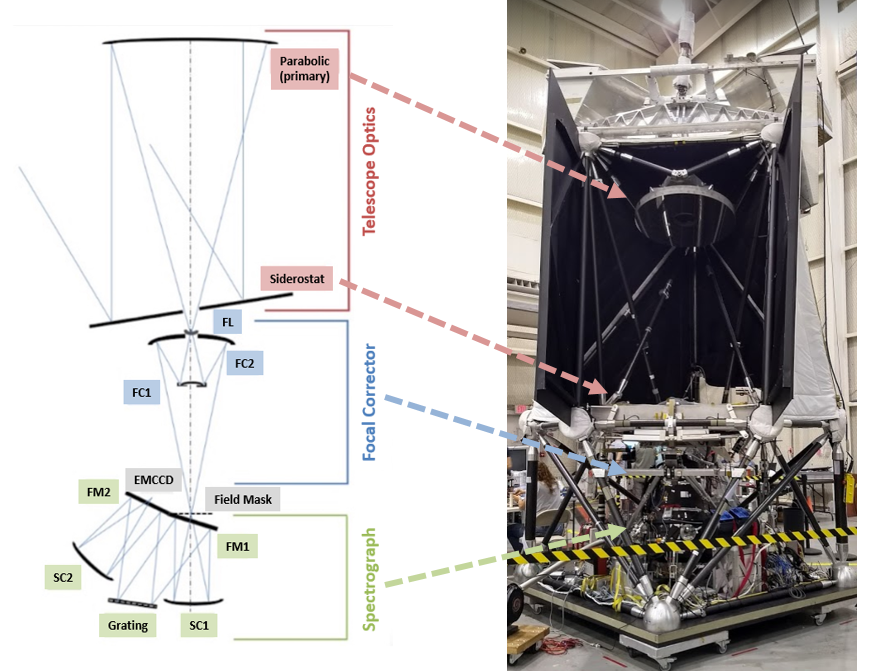}
    \caption{\emph{Left: }The FIREBall-2 light path through the entire instrument assembly. \emph{Right:} FIREBall-2 as-built, awaiting launch during the 2017 Ft. Sumner, NM campaign.} 
    \label{fig:telescope}
\end{figure}

\section{FIREBall-2 design and implementation}

The FIREBall-2 telescope and gondola are the same structures described in \cite{2008Tuttle,2010Tuttle} and are shown in Figure~\ref{fig:telescope}. The telescope assembly consists of a flat 1.2-meter siderostat mirror, which provides elevation control between 40-70$^{\circ}$ and course tip/tilt pointing. This feeds the primary mirror, an f/2.5 1-meter parabolic mirror. Both optics have survived two previous descents and landings (FB-1): the parabola debonded during the first landing, but was otherwise unharmed. Both mirror coatings were stripped by James Mulherin, who originally fabricated them, and re-coated at Goddard Space Flight Center with a standard Al/MgF$_2$ coating optimized for 205 nm. The optics suffered more severe damage after the 2018 flight, and the team is currently having them repaired and re-coating for a Summer 2020 campaign.

\begin{figure}
    \centering
   \includegraphics[width=0.5\textwidth]{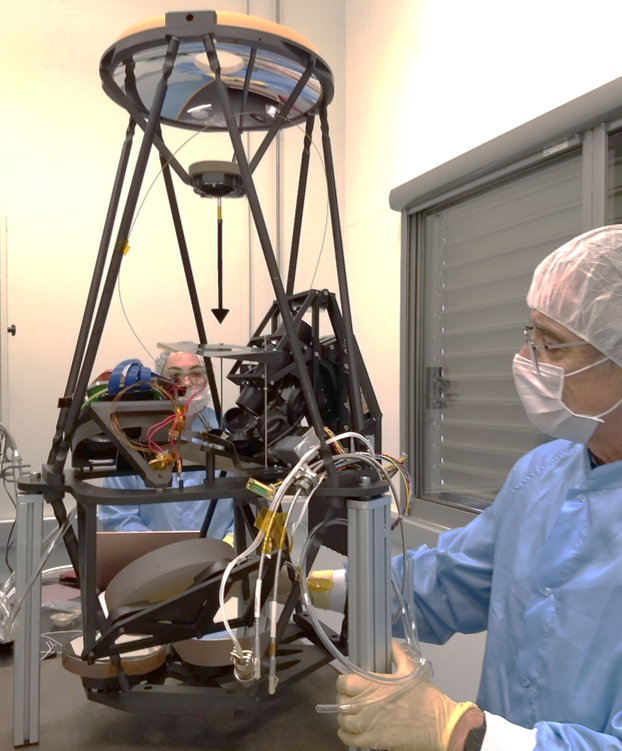}
    \caption{The FIREBall-2 spectrograph with members of the team, Erika Hamden and Bruno Milliard.} 
    \label{fig:spectro}
    \vspace{-0.2cm}
\end{figure}

The FIREBall-2 spectrograph and as-built structures are shown in Figure~\ref{fig:spectro}. The motivation of the optical re-design includes a two-mirror field corrector, which increases the quality of the field over a 35$^{\prime}$ field-of-view. The spectrograph consists of two Schmidt mirrors (one works as a collimator, while the other works as a camera) for compactness. 

\subsection{Technology Demonstrations}

FIREBall-2 acts as a technology demonstration for delta-doping for UV sensitivity, the use of an EMCCD on a sub-orbital payload, and a new grating.

This is a novel, high-throughput, cost-effective anamorphically-aspherized reflective Schmidt grating, manufactured using a double replication process at JY. The 110x130mm grating has a line density of 2400 lines/mm and angle of incidence of 28$^{\circ}$ \cite{2014Grange}. The grating corrects for spherical aberrations and consists of an aluminum substrate with native oxide, followed by a 70 nm layer of MgF$_2$ optimized for FIREBall-2. The grating reflectance exceeds 50\% in band, a significant improvement over the FB-1 grating performance (17\%). The resolving power of the grating is $\sim$2100 at the detector, meeting the FIREBall-2 spectral resolution specified. 

The delta-doped EMCCD is described in Section \ref{sec:EMCCD}.

\subsection{FIREBall-2 observing strategy}

The scientific objectives of FIREBall-2 and the multi-object nature of the spectrograph impose stringent requirements on the instrument pointing. Each science target must be centered in a 50-75 micron slit, corresponding to an accuracy of $\pm$1$^{\prime \prime}$.  In addition, FIREBall-2 aims to detect very faint emission, requiring long integration times ($\sim$50 seconds). During each integration, the instrument must maintain stability to within a pixel on the detector ($\sim$1$^{\prime \prime}$) to avoid degrading the image. The purpose of the CNES designed guidance system is to refine the pre-compensation pointing achieved by the attitude control system. The guidance system corrects residual errors in the pointing by making on-sky measurements of the instrument pointing offset and providing high-cadence feedback to the attitude control system.  

The FIREBall-2 targets were selected using several metrics. The primary UV emission lines (Ly-$\alpha$, OVI, CIV) must fall into the FIREBall-2 bandpass, limiting redshifts to $\sim$ 0.7, 1.0, and 0.3 (respectively). We targeted fields of large scale structure where many sources would be available over the FIREBall-2 FOV. Galaxies at the correct redshift for Ly-$\alpha$ were prioritized, with other emission lines used to fill in spaces in the slit mask as needed. Within Ly-$\alpha$, galaxies were selected based on UV brightness and to provide a variety of galaxy types for each target field. For the 2018 Ft. Sumner campaign, four MOS slit masks were made, with nearly 300 galaxies total targeted.

\subsection{Flight of FIREBall-2}

\begin{wrapfigure}{r}{0.5\textwidth}
\vspace{-1.4cm}
    \includegraphics[width=0.5\textwidth]{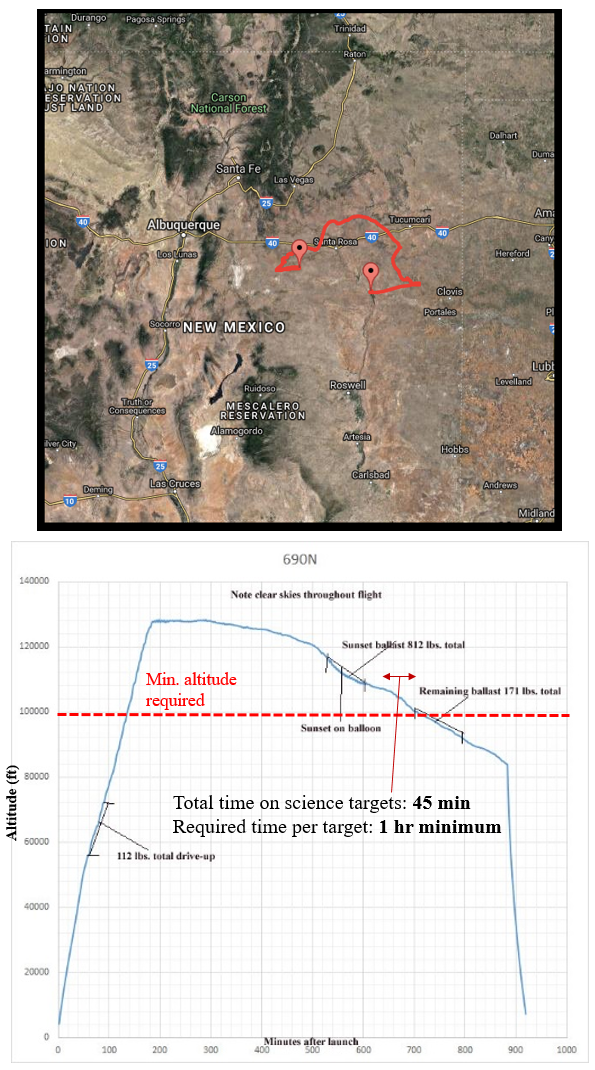}
    \caption{CSBF-generated FIREBall-2 altitude versus time since launch. We note the minimum altitude requirement for FIREBall-2 with a dashed red line. FIREBall-2 was not able to observe its science targets for the required time at night before the balloon fell below its UV transmission window.} 
    \label{fig:flightmap}
    \vspace{-0.4cm}
\end{wrapfigure}

FIREBall-2 was launched from a 40 MCF balloon from Fort Sumner, NM on the morning of September 22nd, 2018 at 10:20 AM local MDT. At launch, all flight systems were nominal and the payload experienced minimal jerks or acceleration during the launch. Throughout the ascent phase (from the surface to float altitude of 128,000 ft), the payload systems and communication continued to behave normally. The payload reached float altitude at approximately 1 PM local time, northeast of the launch site.

The stratospheric winds were very low during the flight and so the balloon traveled slightly north towards Santa Rosa, NM, before eventually drifting south west and ended up over Vaughn, NM. The GPS map of the flight is shown in Figure \ref{fig:flightmap}. The altitude of the payload over the course of the flight is also shown, along with ballast drops and other events. The balloon is expected to lose altitude as the sun sets due to changing thermal conditions and then stabilize at a lower altitude after sunset. Ballast is dropped during sunset to minimize the altitude loss, and the expected nighttime altitude was 118 kft to 108 kft, depending on the particulars of the atmosphere and weather. The balloon lost altitude starting mid-afternoon and continuing through the night. The most likely cause of this was a hole, allowing helium gas to escape. Ballast drops did not help to slow the descent and additionally created some extra pointing challenges since the timing excited a 5 minute resonant frequency in the balloon altitude (buoyancy). The balloon altitude was below the required science minimum altitude of 105,000 ft at 660 minutes after launch (roughly 9:20 PM MDT). We continued to collect data throughout the remainder of the flight, until required to turn off all systems by CSBF prior to termination. The balloon and payload were separated at roughly 1 AM MST, taking about an hour to land. 

The flight was terminated roughly 50 miles west of Fort Sumner, near the municipality of Vaugn, NM. The payload was recovered on September 23rd. The landing, due in part to higher winds at night, was rough and both large optics sustained edge fractures. The siderostat fracture was minimal, while the primary mirror lost roughly 5\% of the area, including the location of two of the 6 bond pads. The damage for both mirrors was on the same side, so was likely a result of a hard landing. In addition, the large mirror of the field corrector has a crack at the edge of the mirror along one of the bond locations. Additional inspection of the remaining mirrors is being conducted. A measurement of the surface figure of both large optics indicates that despite the damage, the surface shapes were not changed significantly and there are no additional cracks or fractures that could propagate through the optic. Both optics will need to have the broken edges sanded down and be re-aluminized before a re-flight.

\subsection{Flight test of delta-doped EMCCDs}\label{sec:EMCCD}

The electron multiplying CCD used on this flight was an Teledyne-e2v CCD201-20 architecture, which consists of 13.5 $\mu$m square pixels, a 2kx1k image area, and an additional 1k pixels added to the serial register. Out of these additional serial register pixels, 604 act as multiplication pixels. When these pixels are clocked with a voltage above $\sim$39 V, they are deep enough to allow impact ionization of electrons as they are moved through the serial register. The net result of this is single electron events in the image area can be read out as 1000 electron events at the amplified, thus significantly reducing the effect of read noise on the resulting signal to noise. A more detailed description of the operation of these devices can be found in Kyne, 2016\cite{2016Kyne}.

\begin{figure}
\vspace{-0.2cm}
\centering
    \includegraphics[width=\textwidth]{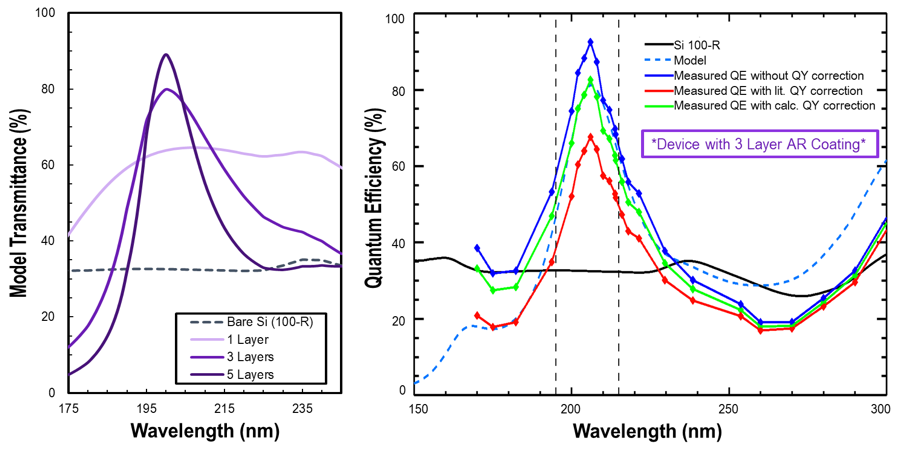}
    \caption{Left: Model transmittance/performance for 2D-doped silicon detectors with multilayer AR coatings tailored for the FIREBall bandpass. Adding complexity increases peak QE, but results in a narrower peak. Theoretical response for a bare 2D-doped silicon detector is also shown. Left: Experimental results for 2D-doped silicon detector with the 3-layer FIREBall AR coating. As measured QE is shown alongside QE corrected for quantum yield.} 
    \label{fig:QE}
    \vspace{-0.2cm}
\end{figure}

The flight device was the end result of several years of technology development undertaken by JPL, Caltech, and Columbia University. In brief, device wafers are bonded to a carrier wafer and back thinned to the epitaxial layer. After thinning, a custom JPL process known as delta-doping is performed, which uses a molecular beam epitaxy (MBE) machine to grow a single layer of boron-doped silicon on the thinned backside \cite{2012Nikzad}. A 2-3nm cap of Si is then grown. The delta-doping process modifies the electric field at the back surface to eliminate the well that typically captures UV photons, which have a very low absorption depth. This brings the internal quantum efficiency (QE) of the device up to 100\%, with all photo-electrons created then detected. The reflection of Si in the UV limits the overall QE of the detectors, and so a custom 3-layer anti-reflection coating was developed to reduce reflection in the FIREBall bandpass \cite{2012Hamden,2014Hamden,2015Hamden,2015Hennessy,2015Jewell}. Measured QE is shown in Figure \ref{fig:QE}

The detector performance was verified via testing at JPL and at Caltech. JPL testing included QE verification \cite{2010J}, while the Caltech test set-up measured QE at limited wavelengths using NUVU v2 and v3 CCCP controllers \cite{2008Daigle} and a custom flight printed circuit board (PCB). These controllers provide 10 ns waveform control to optimize the pixel clocking strategy, readout speed, and wave form shape/height to minimize clock-induced-charge (CIC), read noise, and deferred charge \cite{2015Hamden,2016Kyne}. The PCB was designed using the suggested configuration from NUVU to reduce additional read noise from the readout process. Total read noise in the camera system was 100 e$^-$, which was significant. Measured CIC was 4 $\times$ 10$^{-4}$ e-/pix/frame in the serial register and 6 $\times$ 10$^{-4}$ e-/pix/frame in the parallel clocking, for a total of 1 $\times$ 10$^{-3}$ e-/pix/frame. The dark current level was similar to that of an engineering grade version of the same device, provided by Teledyne-e2v, with 5 $\times$ 10$^{-2}$ e-/pix/hr.

The detector was installed into the payload in the spring of 2016, and was extensively tested in the spectrograph system. Both the detector and NUVU controller performed reliably and with extreme stability thought the integration and testing, showing no change in behavior between installation in 2015 and flight in 2018. 

During the flight, the detector operated in both normal mode and multiplying mode. The photon counting capability was not used in flight due to an excess of scattered light that resulted in higher than expected count rates on the detector ($>$1 event/pix/frame). This is above the typical 0.1 event/pix/frame confusion limit \cite{2008Daigle,2010Daigle,2012Daigle}, which is typically the standard for photon counting. 

The UV sensitivity provided by the delta-doping process worked flawlessly across the FIREBall bandpass (197-213 nm). The science team is still analyzing the data collected during the shortened flight, but we were able to easily measure a spectrum from a GALEX UV bright star, which was observed as a standard. 

This flight test, using delta-doped AR coated EMCCD with a NUVU controller \cite{2008Daigle}, was the first use of UV multi-object spectrograph, and the second time a delta-doped device has flown (the first time was on a rocket payload). 

\section{FIREBall-2 as a pathfinder}

FIREBall-2 acts as a pathfinder both scientifically and technologically. It has already flight-tested both the EMCCD architecture and the delta-doping UV capability, as well as testing out a novel, high throughput, cost-effective, holographic, aspherized reflective grating \cite{2014Grange}.

The science FIREBall-2 is attempting to perform is incredibly ambitious. Measurements of the CGM at higher redshift have been conducted most effectively on 8-m class or larger telescopes (Keck and the VLT) and have really only borne fruit in the last five years. The original concept of FIREBall, which was initially proposed in the early 2000s, was a huge leap given both the knowledge and technology available at the time but has proven to be prescient. The growth of CGM observations via COS and ground based IFUs was still over a decade away when FIREBall was first proposed. The balloon program was able to take a risk on previously unexplored science, and help foster the development of an incredibly important field of galaxy evolution.

\subsection{High Efficiency UV detectors as mission enabling technology}

Photon-counting CCDs are the future for both imaging and spectroscopy in the UV, where the sky background is quite dark \cite{1998Leinert}. The exact nature of that counting, either EMCCDs or farther in the future with Skipper CCDs and MKIDs, is still up for grabs. For a typical IGM emission application, an observation with S/N=6 obtained with a photon-counting CCD detector would drop to S/N=1.4 with a micro channel plate detector and S/N=0.4 for a UV CCD with 2 e- read noise. Like all CCDs, these detectors are also straightforward to produce in large numbers and in a range of formats (up to 4k$\times$4k currently), and would be suitable in mosaics for future large spectrometers. 

Photon-counting CCDs have been based-lined for future large missions such as LUVOIR, HABEX, and WFIRST. Additionally, the controller for these CCDs is in some ways just as important as the CCD itself. The low noise capabilities can only be leveraged with carefully shaped waveforms and pixel clocking speeds. The NuVu v2 controller flew in the 2018 FIRBall-2 flight, and we are planning to further test the NuVu v3 controller for the 2020 flight of FIREBall-2. This exact controller will be used for WFIRST CCD readout, and the sub-orbital program can provide an important flight test of its performance.

Future medium and large UV spectroscopic and imaging missions will likely target the IGM in both absorption and emission, as well as numerous other UV science cases which take advantage of the low sky backgrounds and multitude of processes that have UV signatures. As such, a factor of 10-20 times improvement in detector efficiency corresponds to 3-4.5-fold reduction in telescope size for the same mission performance. Alternatively, when considering future flagship missions, these highly sensitive detectors will improve science payout by an order of magnitude, revealing new aspects of our universe and providing answers to questions we don't even know to ask.

\section{NASA's suborbital balloon program}

These advances in detector technology and eventually in scientific discovery are made possible because of NASA's suborbital program. While operating a balloon payload presents many challenges for the teams involved, not the least of which is weather driven, the sub-orbital program provides a clear path forward for technology advancement. The addition of CubeSats and SmallSats to NASA's portfolio will be a similarly useful addition to the technology lifecycle. 

The advantages of this path for technology development are also clear over all of SMD. It is crucial that as NASA prioritizes large missions, the utility of smaller missions is not forgotten. The suborbital rocket and balloon program are essential for technology development, for trying out innovative, path-finding science, and for training the next generation of instrument builders and PIs. The sub-orbital program not only provides an avenue for technology advancement but it is a critical pipeline to train the future PIs of astrophysics.

\acknowledgments 
 
This work was funded primarily through the APRA program for sub-orbital missions which funds the US side of FIREBall. CNES and CNRS provided support for the French side of the FIREBall collaboration. Additional support for detector development was provided by the SAT program and the Keck Institute for Space Studies. Prof. Hamden's work is partly supported by a Nancy Grace Roman Fellowship, as well as an NSF AAPF and Milliken Prize Fellowship from Caltech during her postdoc. Dr. Hoadley is supported through a Lee Prize Fellowship from Caltech. 
 

\bibliography{report} 
\bibliographystyle{spiebib} 

\end{document}